*In the previous version an error was discovered in the calculation of residual stresses, and also in the use of silica substrates. This is the corrected version, where the conclusions have changed.*

# Elastic Modulus of Polycrystalline Halide Perovskite Thin Films on Substrates

Madhuja Layek, Anush Ranka, In Seok Yang, Zhenghong Dai, Truong Cai, Brian W. Sheldon, Eric Chason, and Nitin P. Padture [*]

School of Engineering, Brown University, Providence, RI 02912, USA

[*]Email: nitin_padture@brown.edu

## ABSTRACT

Using an innovative combination of multibeam-optical stress-sensor (MOSS) curvature and X-ray diffraction (XRD) techniques, the Young's modulus ($E$) of polycrystalline MAPbI$_3$ metal-halide perovskite (MHP) thin films attached to Si substrates is estimated to be 10.2±3.4 GPa. This is comparable to the $E$ of corresponding MAPbI$_3$ single-crystals. This generic method could be applied to other systems to estimate hard-to-measure $E$ of thin films.

## GRAPHICAL ABSTRACT

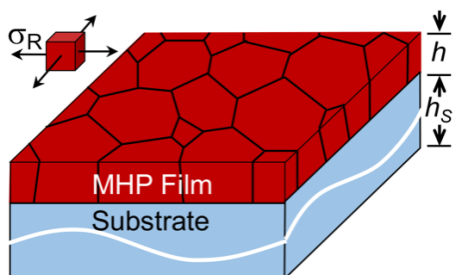

*Estimation of Young's Modulus (E) of Polycrystalline MHP (MAPbI$_3$) Thin Film on Si Substrate*

Tremendous progress has been made in enhancing the efficiency, scalability, and operational-stability of solar cells based on metal-halide perovskites (MHPs).[1] Recently, a positive correlation between the operational-stability of perovskite solar cells (PSCs) and mechanical adhesion toughness of PSC interfaces has been demonstrated.[2, 3, 4] Thus, progress is also needed in the area of mechanical reliability, a key attribute that contributes towards the overall durability of PSCs, before they can be successfully commercialized.[5] Unfortunately, the most basic of mechanical properties, such as the Young's modulus ($E$) (stiffness) of MHP thin films within PSCs, are not known reliably. Methylammonium lead triiodide (CH$_3$NH$_3$PbI$_3$ or MAPbI$_3$), a 3D MHP, is the most studied by far in this regard, and the summary of its $E$ measurements are listed in Table S1 in Supplementary Information (SI). The thin-film $E$ values measured using techniques such as uniaxial tensile testing (free-standing thin films) [6] and nanoindentation [7] range widely, from 5.54 to 23 GPa. The single-crystal $E$ values measured using uniaxial compression testing,[8] nanoindentation,[9] and neutron scattering [10] also range widely, from 10.4 to 23.92 GPa. The reasons for these wide ranges are manyfold.[11] Here we have measured the $E$ of polycrystalline MAPbI$_3$ thin films using a new method which takes advantage of the equi-biaxial tensile residual stress ($\sigma_R$) present in these thin films when they are attached to Si substrates. The primary source of $\sigma_R$ is the coefficient of thermal expansion (CTE) mismatch ($\Delta\alpha$) between the MHP thin film and the substrate, which is given by:[12-14]

$$\sigma_R = \frac{E\Delta\alpha\Delta T}{(1-\nu)}, \qquad (Eqn.\ 1)$$

where $E$ and $\nu$ is the thin-film Young's modulus and Poisson's ratio, respectively, and $\Delta T$ is the difference between the heat-treatment and the use temperatures. Here we have used an innovative combination of multi-beam optical stress sensor (MOSS) curvature (Fig. 1a)[14] and X-ray diffraction (XRD) $\sin^2\psi$ (Fig. 1b)[15] techniques to measure $E$. We take advantage of the fact that the MOSS technique measures residual stress ($\sigma_R$) in the thin film, which does not require *a priori* knowledge of the thin-film $E$. And the XRD $\sin^2\psi$ technique measures residual strain ($\varepsilon_R = \Delta\alpha\Delta T/(1-\nu)$) in the same thin films. Hooke's elasticity law, $E = \sigma_R/\varepsilon_R$, is then used to estimate thin-film $E$, assuming in-plane biaxial isotropy.

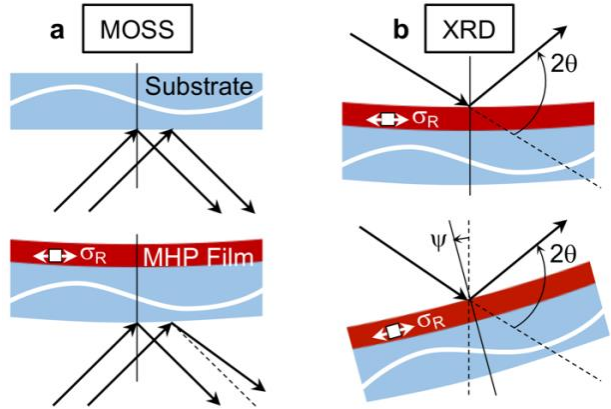

**Figure 1.** Schematic illustrations of:[5] (a) MOSS and (b) XRD $\sin^2\psi$ techniques.

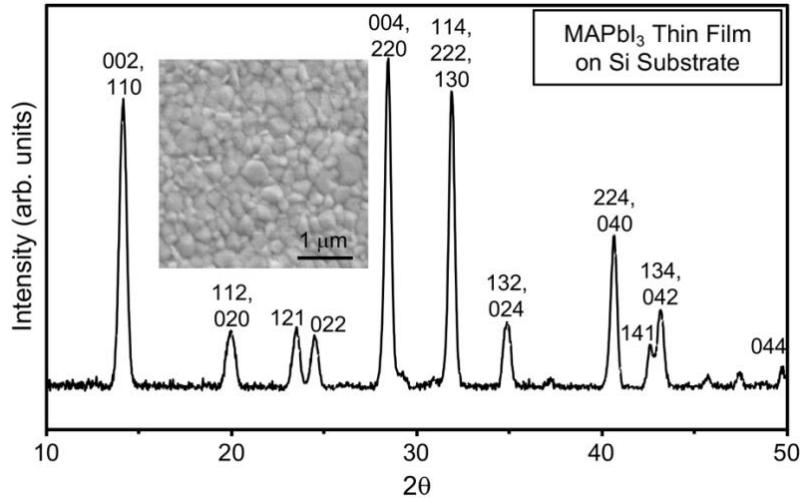

**Figure 2.** XRD pattern from top surface of a MAPbI$_3$ thin film deposited on Si substrate. Inset: top-surface SEM image of a MAPbI$_3$ thin film.

High-quality MAPbI3 thin films were deposited on polished top surfaces of Si substrates, using procedures described in the SI. The average thin-film thickness is 475-500 nm. The bare backside surfaces of the Si substrates are also polished and reflective. Figure 2 presents top-surface XRD patterns of the thin films confirming pure perovskite-phase MAPbI3. The inset in Fig. 2 is top-surface SEM image of the thin films showing a uniform, dense microstructure with an average grain size of ~225 nm. Reference curvatures of the substrates from the bottom side were measured using the MOSS technique, before and after thin-film deposition, using the procedure described in the SI. The residual strain in the very same thin films were measured using the XRD $\sin^2\psi$ technique (see SI).

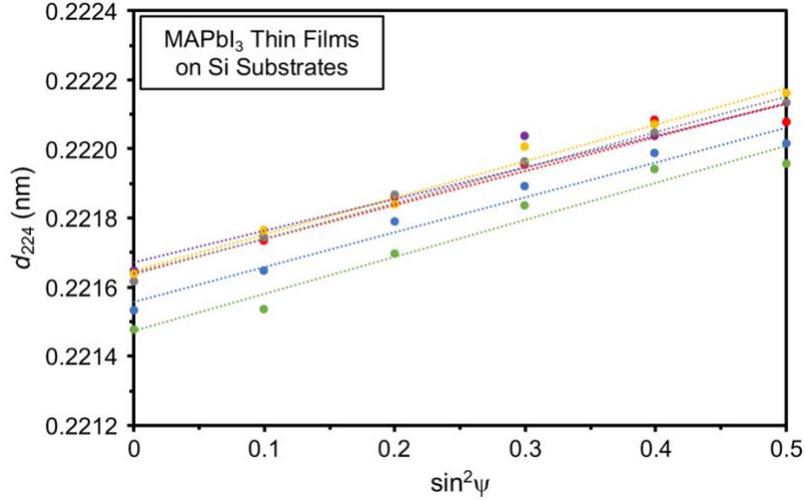

**Figure 3.** XRD $\sin^2\psi$ measurements from MAPbI3 thin films on Si substrates (six specimens). Dashed lines are linear fits.

The $\sigma_R$ in the thin films from the steady-state MOSS curvature measurements were calculated using the Stoney equation:[15]

$$\sigma_R = \frac{E_S h_S^2}{6(1-\nu_S)h}\left(\frac{1}{R_C} - \frac{1}{R_O}\right), \quad \text{(Eqn. 2)}$$

where $E_S$ is the substrate Young's modulus, $\nu_S$ is the substrate Poisson's ratio, $h_S$ is the substrate thickness, $h$ is the film thickness, and $R_O$ and $R_C$ are measured radii of curvature before and after thin-film deposition, respectively. The equi-biaxial tensile $\sigma_R$ is estimated at 34.6±11.5 MPa (average and standard deviation of six specimens). This average $\sigma_R$ are put into the following XRD $\sin^2\psi$ relation to calculate $E_F$ of the thin film:[16]

$$E = (1+\nu)\sigma_R\left(\frac{d_o}{n}\right), \quad \text{(Eqn. 3)}$$

where $n$ is the slope of fitted line to the interplanar $d$-spacing vs. $\sin^2\psi$ data, and $d_O$ is the y-intercept. Figure 3 presents $d$-spacing vs. $\sin^2\psi$ data (linear fits) for MAPbI3 thin films on Si substrate (six specimens). The estimated $E$ value using Eqn. 3 of MAPbI3 thin films on Si substrate is found to be 10.2±3.4 GPa. This $E$ value, which is most relevant to PSCs because it is measured for polycrystalline thin films attached to a substrate, is comparable to single-crystal $E$ of 10.6±1.0 GPa measured in our laboratories using uniaxial compression.[8] (Note that MOSS and XRD $\sin^2\psi$ techniques were utilized by Rolston, *et al.*[13] to measure $\sigma_R$ in mixed-cation-halide MHP thin films.

However, the MOSS and the XRD $\sin^2\psi$ measurements were for thin films on Si and glass substrates, respectively. Also, no effort was made to measure $E$ from those data in that paper.)

According to Eqn. 1, the predicted $\sigma_R$ in MAPbI$_3$ thin film, with a CTE of ~50×10$^{-6}$ °C$^{-1}$ (ref. 17) on Si, with a CTE of ~2.6×10$^{-6}$ °C$^{-1}$ (ref. 18), for $\Delta T$ of 75 °C is ~54 MPa. This is greater than the measured average $\sigma_R$ of 34.6±11.5 MPa. This is possibly because the underlying assumptions in Eqn. 1 — the thin film fully attaches to the substrate only at the heat-treatment temperature and that there is no stress relaxation during cooling — may not be strictly valid. However, this is not expected to affect the $E$ estimation.

To study the possible relaxation of the MAPbI$_3$ thin films at room temperature, time-dependent $\sigma_R$ and $\varepsilon_R$ measurements using MOSS and XRD $\sin^2\psi$, respectively, were performed, as described in the SI. Figure 4 present these data for four different specimens. While the MOSS measurements are continuous (after initial 2-3 minutes delay), each XRD $\sin^2\psi$ measurement takes ~12 min. The MOSS results show initial rapid (~12 min), but modest, relaxation in $\sigma_R$ of 4-5 MPa. No appreciable relaxation in $\varepsilon_R$ is observed because the XRD $\sin^2\psi$ measurement duration is relatively long. Since MOSS measurements were performed after >12 min, the $E$ estimation above is not expected to be affected by the initial rapid stress relaxation.

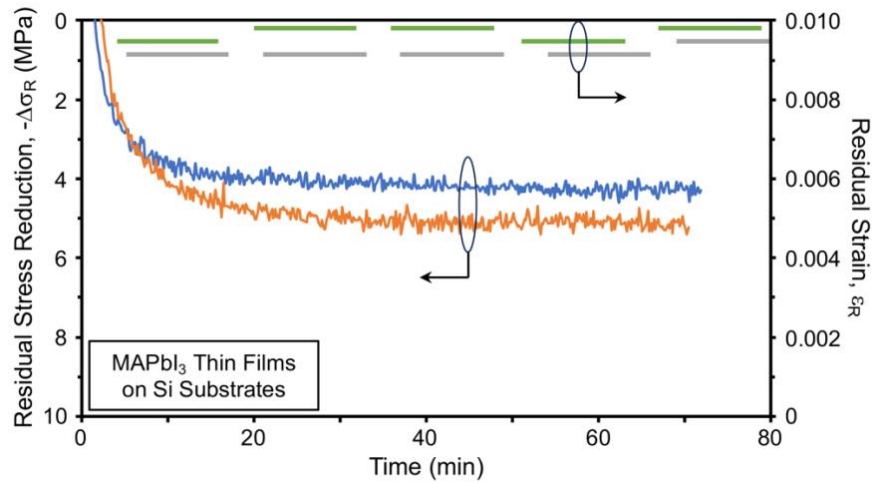

**Figure 4.** Time-dependent $\sigma_R$ and $\varepsilon_R$ measurements using MOSS and XRD $\sin^2\psi$, respectively, in MAPbI$_3$ thin films on Si substrates (two different specimens for $\sigma_R$ and $\varepsilon_R$ measurements each).

In summary, using a combination of MOSS and XRD techniques, the $E$ of polycrystalline MAPbI$_3$ thin films attached to Si substrates is estimated to be 10.2±3.4 GPa, which is comparable to that of corresponding $E$ of MAPbI$_3$ single-crystals. This generic method could be applied to other systems to estimate hard-to-measure $E$ of thin films.

## ACKNOWLEDGEMENTS

Funding for this work was provided by the Office of Naval Research (Grant No. N00014-20-1-2574) and the National Science Foundation (Grant No. 2102210).

Methylammonium Lead Perovskites Using X-ray Diffraction. *Inorg. Chem.* **54**, 10678 (2015).
18. Bartl, G.; Elster, C.; Martin, J.; Schödel, R.; Voigt, M.; Walkov, A.: Thermal Expansion and Compressibility of Single-Crystal Silicon Between 285 K and 320 K. *Measur. Sci. Technol.* **31**, 065013 (2020).

# SUPPLEMENTARY INFORMATION
**Supplementary Table**

**Table S1.** MAPbI$_3$ Young's modulus ($E$) experimental values reported in the literature.

| $E$ (GPa) | Measurement Technique | Form | Reference |
|---|---|---|---|
| 10.2±3.4 | XRD + MOSS | Thin Film | This work |
| 13.5 | Nanoindentation | Thin Film | S5 |
| 19.65±2.45 | Nanoindentation | Thin Film | S6 |
| 23 | Nanoindentation | Thin Film | S7 |
| 8.6 - 24.6 | Nanoindentation | Thin Film | S8 |
| 11.3 - 18.8 | Nanoindentation | Thin Film | S9 |
| 16.5±2.0 | Nanoindentation | Thin Film | S10 |
| 17.7±1.9 | Nanoindentation | Thin Film | S11 |
| 22.8±3.1 | Nanoindentation | Thin Film | S12 |
| 10.9 | Atomic Force Microscopy | Thin Film | S13 |
| 5.54±0.73 | Direct Uniaxial Tension | Free-Standing Thin Film | S14 |
| 10.4±0.8 | Nanoindentation | Single-Crystal | S15 |
| 14.3±1.7 | Nanoindentation | Single-Crystal | S16 |
| 10.8±2.7 | Nanoindentation | Single-Crystal | S17 |
| 20.0±1.5 | Nanoindentation | Single-Crystal | S18 |
| 12.7 - 17.8 | Nanoindentation | Single-Crystal | S19 |
| 23.92±3.63 | Nanoindentation | Single-Crystal | S20 |
| 12.6 | Nanoindentation | Single-Crystal | S21 |
| 12 | Nanoindentation | Single-Crystal | S22 |
| 18.9 | Atomic Force Microscopy | Single-Crystal | S21 |
| 14.1 | Neutron Scattering | Single-Crystal | S23 |
| 10.6±1.0 | Direct Uniaxial Compression | Single-Crystal | S24 |

## Experimental Procedure

*Materials and Thin Film Preparation*

MAPbI$_3$ precursor solution was prepared by dissolving methylammonium iodide (MAI; Greatcell, Australia) and Pb(II) iodide (PbI$_2$; 99.99%, TCI) in stoichiometric (1:1) ratio in a mixed solvent of 9:1 volume ratio of N,N-dimethylformamide (DMF; 99.8%, Sigma-Aldrich, USA) and dimethyl sulfoxide (DMSO; 99.7%, Sigma-Aldrich, USA) in a N$_2$-filled glovebox. Double-side polished Si <100> wafers with 0.2 mm thickness and 25.4 mm diameter (University Wafer, USA) were used as substrates. The polished surfaces had thermally grown oxide (~300 nm thickness) on each side. The substrates were cleaned sequentially in ultrasonic baths of detergent solution, DI water, acetone, and isopropanol. They were subsequently UV-ozone treated for 20 min before thin-film deposition. The MAPbI$_3$ thin films (475-500 nm thickness) were deposited on the pre-cleaned substrates by spin-coating the precursor solution at 4,000 rpm for 20 s. Diethyl ether (Sigma Aldrich, USA) antisolvent (250 μL) was dripped at the center of the film after 10 s of spinning. The as-deposited films were then annealed at 100 °C for 10 min on a hot-plate. All the thin films were prepared in a humidity controlled dry-box (11% RH).

*Characterization*

The top-surfaces of the MAPbI$_3$ thin films were observed using a scanning electron microscope (SEM; Quattro ESEM, ThermoFisher Scientific, USA). The linear-intercept method was used on representative SEM micrographs to determine the grain sizes of the MAPbI$_3$ thin films. The naturally formed grooves between grains were assumed to represent the grain boundaries.

X-ray diffraction (XRD) was performed to confirm the phase purity of all the MAPbI$_3$ thin using a 2D diffractometer (Cu K$_\alpha$ radiation; Discovery D8, Bruker, Germany) in air.

*Mechanical Properties Measurement*

XRD sin$^2\psi$: Using the well-established and internally-consistent XRD sin$^2\psi$ method the equi-biaxial tensile strains ($\varepsilon_R$) in the MAPbI$_3$ thin films were determined using six different specimens in air. The (224) Bragg reflection in the XRD patter was used. The following relation was applied:[S1]

$$\varepsilon_R = \frac{n}{(1+v)d_0}, \quad \text{(Eqn. S1)}$$

where $n$ is the slope of the linear fits to the (224) interplanar spacing $d_{224}$ vs. sin$^2\psi$ data, $d_0$ is $d$ spacing at $\psi = 0$ ($y$-intercept), and $v$ is the MAPbI$_3$ Poisson's ratio. XRD patterns at different $\psi$ angles (0° to 45°) were collected and the systematic shift in the (224) peak position was used to generate the $d_{244}$ vs. sin$^2\psi$ plots. An $v$ value of 0.33 was used.[S2] Typical measurement on one specimen takes ~12 minutes. To study any relaxation in the thin films, the $\varepsilon_R$ values were measured as a function of time on two different specimens.

Multi-Beam Optical Stress Sensor (MOSS): The MOSS (k-Space, USA) technique was employed to measure the residual stress ($\sigma_R$) in the MAPbI$_3$ thin films in air by measuring the change of curvature of the substrate before and after thin-film deposition. Substrate backside reflective surface was used to measure the change in multi-beam spacings due to the change in curvatures. The position and the alignment of the samples were carefully monitored while measuring the change of curvature. The steady-state curvature measurements were preformed ~12 min after the deposition was complete, and average from the first 60 s of curvature-change measurements were used for subsequent calculations. The Stoney equation (Eqn. 2) was used to calculate the thin film stress from the change in film curvature.[S3] Average $E_S$ and $v_S$ values of 150 GPa and 0.17 for Si,[S4] respectively, were used. To study any relaxation in the thin films, the $\sigma_R$ values were measured continuously (~2 minutes after the thin film processing was complete) as a function of time on two different specimens.

**Supplementary References**

S1. Luo, Q.; Jones, A. H., High-Precision Determination of Residual Stress of Polycrystalline Coatings Using Optimised XRD-sin$^2\psi$ technique. *Surf. Coat. Technol.* **2010,** *205,* 1403-1408.

S2. Feng, J., Mechanical Properties of Hybrid Organic-Inorganic CH$_3$NH$_3$BX$_3$ (B = Sn, Pb; X = Br, I) Perovskite Solar Cell Absorbers. *APL Mater.* **2014,** *2,* 081801.